\documentclass[twocolumn,aps,showpacs,preprintnumbers,amsmath,amssymb,ltxgrid,rotate,array,reprint]{revtex4-1}
\usepackage{graphicx,color,epsfig,tabularx,array}
\usepackage{tabularx}
\usepackage{subfig}
\usepackage{multirow}
\begin{document}

\bibliographystyle{apsrev}

\title{Electric field-induced oxygen vacancies in YBa$_2$Cu$_3$O$_7$}

\author{A. Lorenzo Mariano}
\affiliation
{Univ. Grenoble Alpes, CNRS, SIMAP, 38000 Grenoble, France}
\author{Roberta Poloni}
\email{roberta.poloni@grenoble-inp.fr}
\affiliation
{Univ. Grenoble Alpes, CNRS, SIMAP, 38000 Grenoble, France}

\preprint{version {\today}}
\date{\today}
\begin{abstract}
  The microscopic doping mechanism behind the superconductor-to-insulator transition of a thin film of YBa$_2$Cu$_3$O$_7$ was recently identified as due to the migration of O atoms from the CuO chains of the film.  Here we employ density-functional theory calculations to study the evolution of the electronic structure of a slab of YBa$_2$Cu$_3$O$_7$ in presence of oxygen vacancies under the influence of an external electric field.  We find that under massive electric fields isolated O atoms are pulled out of the surface consisting of CuO chains. As vacancies accumulate at the surface, a configuration with vacancies located in the chains inside the slab becomes energetically preferred thus providing a driving force for O migration towards the surface. Regardless of the defect configuration studied, the electric field is always fully screened near the surface thus negligibly affecting diffusion barriers across the film.
  
\end{abstract}
\maketitle

\section{Introduction}
The electric double layers (EDL) technique has been employed during the past 15 years to induce extremely large electric fields at the interface with the oxides \cite{Fuji2013} by making use of ionic liquids as gate dielectrics. This has allowed to achieve extremely high carrier densities that accumulates at the interface to screen the electric field and that in turn modify the electronic structure of the oxide \cite{ShiAsa2007,YeIno2010,LeeCle2011,YamUen2011,Nakano2012}. In the case of high-temperature cuprate superconductors, the EDL technique has been used to efficiently tune the electronic properties of the oxide from the superconducting to the insulating phase \cite{Boll2011,LenBar2012,Javiprb2013,PhysRevB.92.020503}.
In the case of gating experiments performed on VO$_2$-based EDL transistor, the suppression of the metal-to-insulator transition upon electrolyte gating was found to involve oxygen migration from the oxide to the ionic liquid \cite{JeoAet2013,JeoAet2015}. Following this discovery, EDL-gating experiments inducing a superconductor to insulator transition in YBa$_2$Cu$_3$O$_{7-\delta}$ \cite{Leng2011} have been recently revisited \cite{Javipaper}. The change in x-ray absorption spectra collected in situ while measuring transport properties during gating were interpreted using density-functional theory calculations. The spectral changes were associated to a deoxygenation of the sample thus confirming that the doping mechanism behind the change in resistivity was primarly a chemical doping achieved by formation of oxygen vacancies \cite{Javipaper}.

\begin{figure*}[ht]
\includegraphics[scale=0.4]{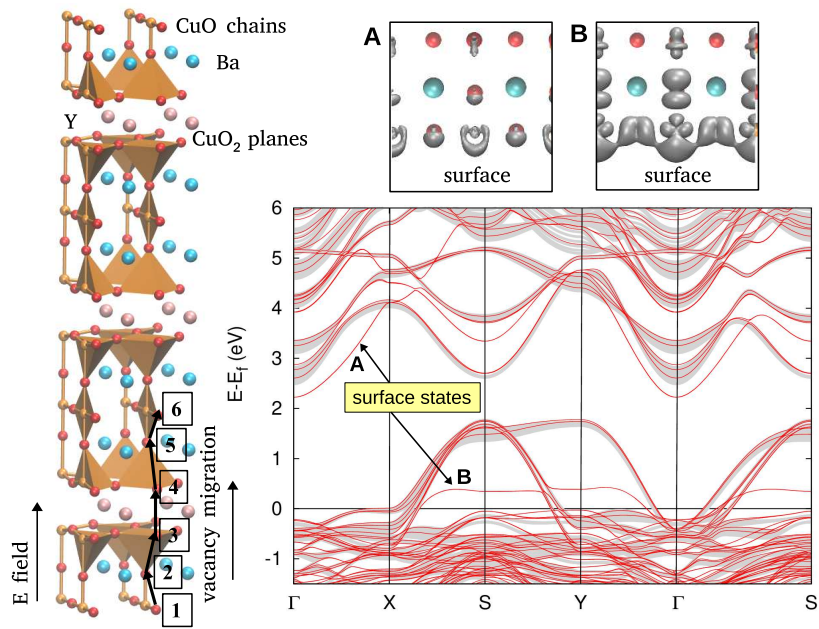}
\begin{centering}
\caption{Left: crystal structure of the slab of YBa$_2$Cu$_3$O$_7$ studied in this work. A 2$\times$2 in-plane supercell is shown here repeated three times along $z$. Cyan and pink atoms represent Ba and Y, respectively. The images and vacancy migration path studied using NEB are also shown. Right: zero-field band structure of the three-unit cells slab (red lines) superimposed with the surface-projected band structure of the corresponding bulk material (shaded bands). The isosurface plots (2$\times$10$^{-2}$ bohr$^{-3}$) of the charge density corresponding to eigenstates (A) and (B) at $\Gamma$ and S, respectively, are also shown.}
\label{fig1}
\end{centering}
\end{figure*}

In a follow-up study \cite{Poloni2018}, we employed density-functional theory (DFT)-based calculations to compute Kohn-Sham-based Cu-K edge x-ray spectra and study the electronic structure resulting from different doping mechanisms. Although we unambigously identified the doping mechansim as the migration of O atoms from the CuO chains (see Figure~\ref{fig1}) \cite{Javipaper,Poloni2018}, many questions still remain concerning the role of the external field on the electronic structure of YBCO and its influence in the vacancy formation process. For example, we wish to clarify the extent and localization of charge doping, its role in the vacancy formation energy and the change of the electronic structure upon O migration.
In order to address these questions we study the evolution of the electronic structure of a thin film of YBCO composed by three or two unit cells along the $z$ direction as a function of external electric fields here silumated by adding self-consistently a saw-tooth potential to the bare ionic potential. Several O-vacancy concentrations and configurations within the film are studied and the computed formation energies for increasing fields (from 2 V/nm to 30 V/nm) are discussed.
The predicted negative formations energies show that massive fields can pull O atoms from the surface, in agreement with low-temperature experimental evidence \cite{Leng2011,Javipaper}. As vacancies accumulate on the surface, the relative stability of surface and $bulk$ vacancies change thus driving O atoms to migrate to the surface. For none of the oxygen-deficient configurations studied here, the applied field can penetrate into the slab. This result together with the weak field-evolution of the vacancy formation energies suggest that the primary role of the field is to create O-vacancy defects on the surface. The large diffusion barriers computed using nudged-elastic band calculations suggest that the simple O migration models employed here cannot describe the dynamics of the process.

\begin{figure*}[ht]
\includegraphics[scale=0.55]{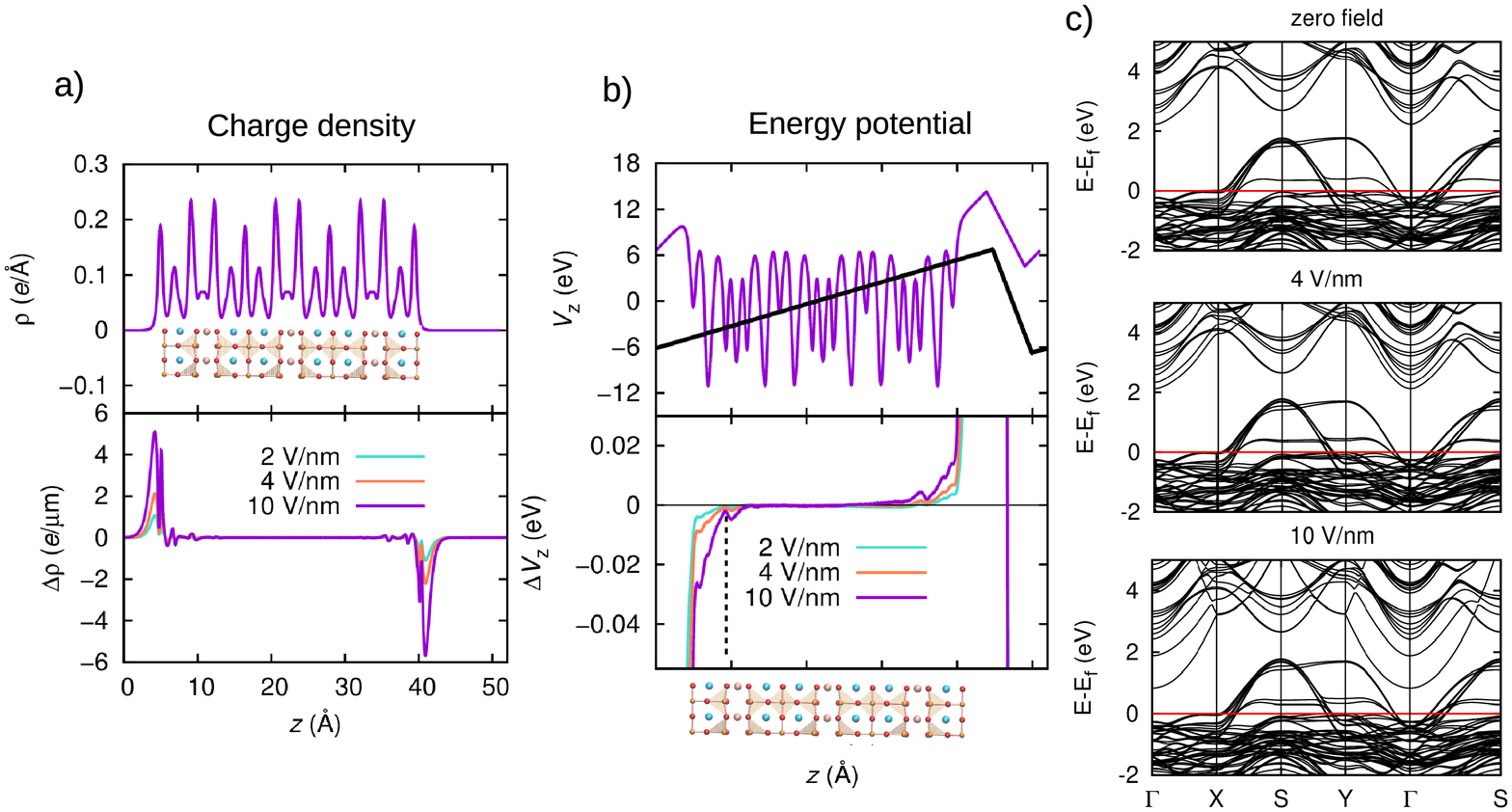}
\begin{centering}
\caption
    {a) Planar average of the valence charge density, $\rho$, as a function of $z$ (upper panel). Difference between the planar average of charge density in presence of electric field and the zero-field case,  $\Delta\rho$ (lower panel). b) Planar average of the electrostatic potential energy for 4 V/nm and sawtooth potential (black line) (upper panel).  Change in electrostatic potential with respect to the zero-field case (lower panel). The vertical dashed line corresponds to the first CuO$_2$ plane from the surface. c) Electronic band dispersion of the three-unit cell slabs computed using an in-plane primitive tetragonal cell (see text). The coordinates of the in-plane high symmetry k-points are X=[1/2,0,0], S=[1/2,1/2,0], and Y=[0,1/2,0].}
\label{fig2}
\end{centering}
\end{figure*}

\section{Computational Details}
The electronic structure and geometrical optimization were performed using the {\sc pwscf} utility of {\sc Quantum espresso} \cite{quantumespresso}. We use the PBE functional, Rabbe-Rabe-Kaxiras-Joannopoulos ultrasoft pseudopotentials \cite{rrkjus-pseudo} available within the pslibrary, and wavefunctions and charge density cutoffs of 50 Ry and 450 Ry, respectively. Geometrical optimizations are performed until the forces are lower than 0.003 eV/\AA\ and the stress along $z$ is less than 0.015 kbar. Calculations were performed using a slab-vacuum geometry: slabs are always CuO chains-terminated to reflect experimental evidence \cite{PhysRevLett.69.2967,PhysRevLett.88.097002} and consistent with a neutrally-terminated surface and estimated lowest surface energy \cite{1997Granozio}.

Because experimentally \cite{Leng2011,Javipaper} YBCO is grown epitaxially on top of PBCO/STO yielding a lattice mismatch with STO of -1.4\% and -0.7\% along the $a$ and $b$ axes, respectively, we adopt here a tetragonal lattice yielding a similarly strained lattice and impose a lattice constant of 3.9419 \AA\ for the primitive cell. 2$\times$2 supercells within the plane are then generated using this lattice constant. Geometrical optimizations are performed by fixing the in-plane lattice constant in order to be consistent with experimental procedure. The Brillouin zone is sampled using a 9$\times$9 and a 5$\times$5 Monkhorst-pack k-point grids for the primitive and 2$\times$2 supercell, respectively, for band structure calculations. For the geometrical optimization of vacancies on the 2$\times$2 supercell, a 3$\times$3 k-point grid is used. In this case, the crystal was repeated only twice along $z$ in order to reduce the computational resources. We nonetheless confirmed that this has a negligible effect on the vacancy formation energy ({\sl vide infra}). The surface-projected bulk band structures and the surface bands of the slabs are plotted by aligning the local self-consistent potential of the bulk with the same potential in the center of the slab.
Calculations are performed by applying the dipole correction \cite{Bengtsson1999,Vander2001}. 
The intrinsic dipole and the external electric field potential are treated self-consistently by adding a sawtooth potential \cite{Resta1986} to the bare ionic potential in the direction normal to the surface such that a net electric field of the desired amount is applied. To recover periodicity, the potential must be discontinuous and we make sure that the discontinuity lies in the middle of the vacuum region. The Climbing Image-Nudged Elastic Band (CI-NEB) calculations for the migration of vacancies through the slab are performed using a 7$\times$7 Monkhorst-pack k-point grid for the primitive cell and a 3$\times$3 grid for the 2$\times$2 supercell. Calculations were converged when forces orthogonal to the path are less than 0.1 eV/\AA.

\section{Results}
\subsection{Electric field applied to stoichiometric YBa$_2$Cu$_3$O$_7$}
We first study the electric field profile of a slab composed by three unit cells along the $z$ direction and primitive within the plane as shown in Figure \ref{fig1}.  This choice is consistent with experiments where also a film composed by three unit cells was grown on top of STO/PBCO \cite{Javipaper}.
%
Figure \ref{fig1} shows the comparison at zero field between the slab band structure (red lines) and the surface-projected bulk bands (shaded grey). A few surface states arising from the termination of the slab are also illustrated as these will be more strongly affected by the applied field.
We apply an electric field of increasing strength to the three-unit cells slab and analyze the response in terms of valence charge density, $\Delta\rho$, and potential energy, $\Delta V_z$. The planar average of the charge density and the electrostatic potential energy along the $z$ direction is shown on the top panels of Figure \ref{fig2} for the slab under zero field.
As expected, electrons accumulate and deplete near the surfaces. However, screening of the field at the interface is not complete and the electric field penetrates a few atomic layers. We remind that YBCO is a 2D metal with metallic states localized mainly within the CuO$_2$ planes and along the apical O sites \cite{Lopez2010,Poloni2018}. A full screening is achieved in correspondance of the first CuO$_2$ metallic plane whose position is indicated (only on one side of the slab for clarity) by the vertical black dashed line in Figure \ref{fig2}. The planar average of  $\Delta V_z$ thus gives a finite value of the dielectric constant within the first three atomic layers. Because no ionic contribution is considered here, this approach yields the optical dielectric constant $\epsilon_{\infty}$. We compute $\epsilon_{\infty}$ as $D$/$E$ where $D$ is the applied electric field in absence of dielectric and $E$ is the planar average of the electrostatic potential, i.e. the screened field \cite{Nakamura2006,Gallifield2010}. Practically, we use the ratio between the coefficient of the linear regression of $D$ and $E$, i.e. between the bare potential outside the slab, in the vacuum region, and the screened potential inside the slab within a few atomic layers. We compute a value of 147. The same result was obtained by using the effective screening medium with the metal-slab-metal by Otani and Sugino \cite{PhysRevB.73.115407}. Because the saw-tooth potential scheme results in a significantly faster convergence of the self-consistent field, we adopt the former method throughout this work.
Consistent with the charge doping occurring only at the surface (Figure~\ref{fig2}, panel a), the changes in the slab band structure in presence of electric field mostly affect the surface states shown in Figure~\ref{fig1} and the most visible effect is the lifting of the degeneracy of these states (see panel c in Figure \ref{fig2}). 

\subsection{Formation energy of oxygen vacancies}
The vacancy formation energy for three O-defect concentrations at varying configurations is computed as a function of the applied field. In our previous study \cite{Poloni2018}, we predicted significantly larger formation energies for vacancies within the CuO$_2$ plane (2.10 eV) as compared to vacancies within the CuO chain (1.20 eV) for YBa$_2$Cu$_3$O$_{6.5}$, assuming all O vacancies aligned in the same chain. This is consistent with previous computational studies \cite{Lopez2010} and the experimental structural characterization of varying YBa$_2$Cu$_3$O$_{7-\delta}$ stoichiometries \cite{Werder1988,Kha1988,Beyers1989}. Thus, in what follows we neglect the study of vacancies located within the CuO$_2$ planes.
\begin{figure}[ht]
\includegraphics[scale=0.30]{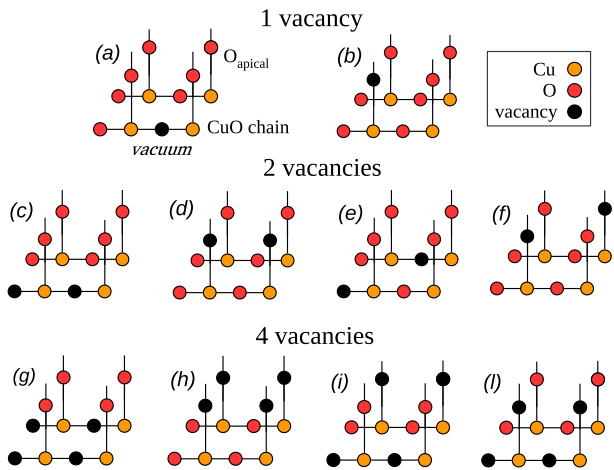}
\begin{centering}
\caption
{Configurations of oxygen vacancies studied in this work using a 2$\times$2 in-plane supercell. The configurations with vacancies located in the chains inside the slab are not shown here and are referred to in the text as $(a')$, $(c')$, $(e')$, and $(g')$.}
\label{schema}
\end{centering}
\end{figure}

We compute O vacancies localized i) within the CuO chain at the surface, ii) within the apical sites near the surface and iii) within the chain $inside$ the slab, i.e. one unit cell below the surface along $z$, and some combinations of i) and ii). Configurations i), ii) and iii) correspond respectively to images 1, 2, and 6, shown in Figure~\ref{fig1}. We compute 1 vacancy, 2 vacancies and 4 vacancies per 2$\times$2 supercell. We take a 2 unit-cells slab along $z$ and perform a full atomic relaxation by fixing the inplane lattice parameter and the atomic position of the CuO termination (away from the vacancies). The configurations of O vacacancies are schematically illustrated in Figure~\ref{schema}. In order to validate the use of a two-unit cell slab along $z$, configurations $(g)$ and $(h)$ were also computed using a cell repeated three times and a negligible difference was found.
The O vacancy formation energy was computed as follows:
\begin{equation}\label{eq1}
\Delta E_{\text{vac}} = \frac{E_{\text{vac}}-E_{\text{stoi}}+n\mu_{O}}{n},
\end{equation}
where E$_{\text{vac}}$ is the total energy of YBa$_2$Cu$_3$O$_{7-\delta}$ optimized slab containing $n$ vacancies, E$_{stoi}$ is the total energy of stoichiometric YBa$_2$Cu$_3$O$_7$, and $\mu_O$ is the chemical potential of O, here taken as half the total energy of the O$_2$ molecule.
\begin{table*}[ht]
  \renewcommand{\arraystretch}{1.3}
  \begin{tabularx}{\textwidth}{XXXX|XXXXXX|XXXXX}
\hline \hline
\multicolumn{15}{c}{$\Delta E_{vac}$ / eV}                                                                                                                                                                    \\ \hline \hline
\multicolumn{1}{l}{\multirow{2}{*}{\begin{tabular}[c]{@{}l@{}}Field \\ V/nm \end{tabular}}} & \multicolumn{3}{c}{1 vacancy} & \multicolumn{6}{c}{2 vacancies}               & \multicolumn{5}{c}{4 vacancies}       \\ \cline{2-15} 
\multicolumn{1}{l}{}                                                                         &   $(a)$      &    $(b)$      & $(a')$    & $(c)$   & $(d)$   & $(c')$  & $(e)$   & $(f)$   & $(e')$  & $(g)$   & $(h)$   & $(g')$  & $(i)$   & $(l)$   \\ \hline \hline
zero                                                                                         & 0.611    & 1.401    & 1.465   & 1.940 & 1.403 & 1.319 & 0.775 & 1.570 & 1.485 & 2.135 & 1.698 & 1.363 & 1.849 & 2.523 \\ \hline
4                                                                                            & 0.556    & 1.426    & 1.461   & 2.042 & 1.428 & 1.318 & 0.710 & 1.606 & 1.495 & 2.196 & 1.737 & 1.362 & 1.906 & 2.541 \\ \hline
6                                                                                            & 0.520    & 1.442    & 1.464   & 2.085 & 1.440 & 1.317 & 0.680 & 1.623 & 1.493 & 2.222 & 1.759 & 1.361 & 1.931 & 2.550 \\ \hline
10                                                                                           & 0.456    & 1.470    & 1.460   & 2.162 & 1.468 & 1.315 & 0.608 & 1.660 & 1.491 & 2.261 & 1.806 & 1.355 & 1.974 & 2.569 \\ \hline  \hline
\end{tabularx}
\caption{Electric field-evolution of $\Delta E_{vac}$ (eV) for the configurations shown in Figure~\ref{schema}. The prime symbol corresponds to configurations of vacancies in the chains located one unit cell inside the slab (see text and caption of Figure~\ref{schema}).}
\label{tab1}
\end{table*}

\begin{figure*}[ht]
\includegraphics[scale=0.57]{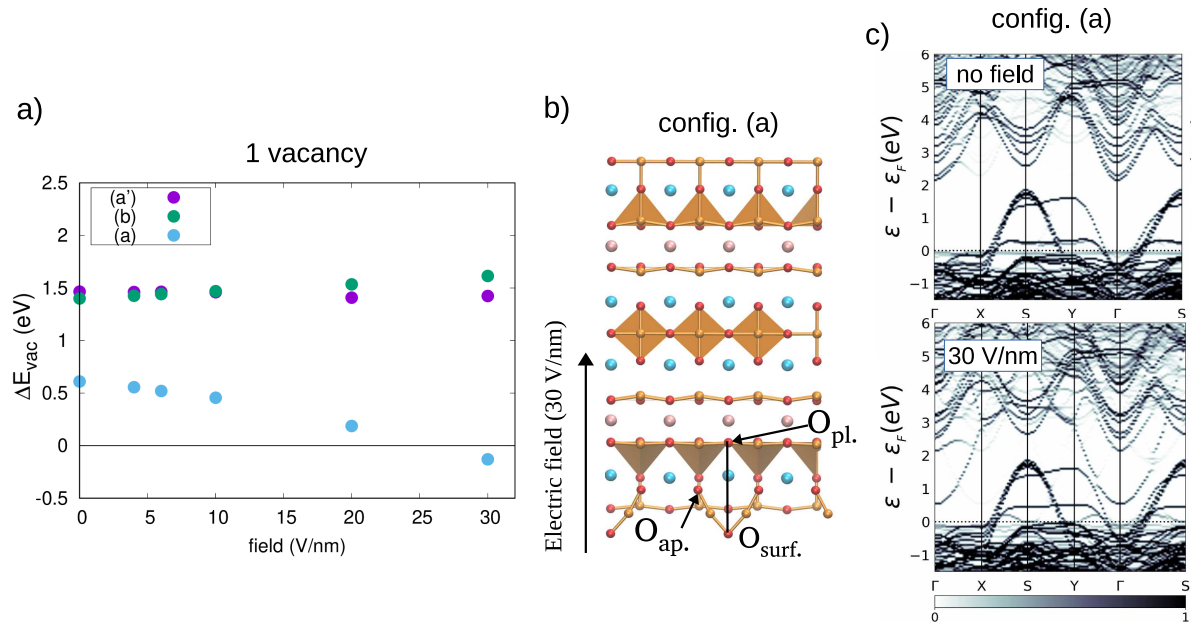}
\begin{centering}
\caption
{a) Evolution of the oxygen-vacancy formation energy computed for configurations $(a)$, $(b)$ and $(a')$ for increasing fields. b) Geometrically optimized slab for $(a)$ at 30 V/nm. c) Unfolded bands of configuration $(a)$: the grey color bar represents the number of bands at a specific k-vector and energy (see Refs.\cite{PhysRevB.89.041407,PhysRevB.91.041116} for more details).}
\label{30V}
\end{centering}
\end{figure*}

For each defect configuration within the chain shown in Figure~\ref{schema} (image 1 in Figure~\ref{fig1}), the corresponding configuration within the chain inside the slab (image 6) is here referred to with the prime symbol. These are not reported in Figure~\ref{schema} for clarity. For example, configuration $(a)$ refers to one vacancy on the CuO chain at the surface and $(a')$ refers to a vacancy in the CuO chain inside the slab. The same applies for $(c)$ and $(c')$, $(e)$ and $(e')$, and $(g)$ and $(g')$.
The O-vacancy fomation energies are reported in Table \ref{tab1}.
The lowest energy of formation is predicted when a single vacancy per supercell is imposed at the surface. i.e. for $(a)$, followed by a similar configuration when two non-aligned vacancies are formed pr unit cell, i.e. configuration $(e)$. The defect formation energy for these configurations significantly decreases as a function of the magnitude of the applied external field while for the rest of configurations the field-dependence is weaker. Thus, we compute the formation energy for $(a)$ even for stronger fields, up to 30 eV/nm, and the results are reported in Figure~\ref{30V}, together with results for configurations $(b)$ and $(a')$ for comparison. We stress that for every calculation we made sure that no electrons are pulled into the vacuum. Remarkably, the formation energy of configuration $(a)$ becomes negative at the most massive field imposed suggesting the spontaneous formation of isolated O vacancies. From the trends shown in Table \ref{tab1}, we expect negative formation energies at even stronger fields for configuration $(e)$. Vacancies at the apical sites in configuration $(b)$ are instead destabilized with increasing field strengths and vacancies in the chain located inside the slab in configuration $(a')$ are instead marginally affected by the field. The geometrically-optimized $(a)$ configuration is shown in panel b of Figure~\ref{30V} and a similar surface geometry with O$_{\text{surface}}$ moving outwards was found for configuration $(e)$. The remaining O atoms in the partially emptied chain move outwards along $z$ and lie within the plane formed by O$_{\text{apical}}$ and the chain Cu. The  O$_{\text{apical}}$-Cu and Cu-O$_{\text{surface}}$ bond lengths are 1.79 and 1.77 \AA\, and the O$_{\text{plane}}$-O$_{\text{surface}}$ is 4.45 \AA. These bonds largely lengthen with the extenal field and become 1.82 \AA\ and 1.85 \AA\ for the Cu-O bonds and the O$_{\text{plane}}$-O$_{\text{surface}}$=6.27 \AA\ at 30 V/nm, as shown in Figure~\ref{30V}, panel b. The band structure of configuration $(a)$ is shown at zero field and 30 V/nm in Figure~\ref{30V}, panel c.  In order to assist the analysis of the electronic structure a band unfolding procedure was used to recover the primitive cell picture of a 2$\times$2 supercell \cite{PhysRevB.89.041407,PhysRevB.91.041116}. The surface states shown in Figure~\ref{fig1} are modified by the presence of defects and are strongly affected by the electric field. The surface state labelled as (B) in Figure~\ref{fig1} and consisting of surface Cu d$_{\text{y$^2$-z$^2$}}$ hybridized with p states of apical and surface O, at massive fields (30 V/nm) and in presence of an O vacancy,  crosses the Fermi level at S and Y (see panel c in Figure~\ref{30V}).

At zero field, in the case of one vacancy per unit cell the lowest formation energy is computed for an O-deficient site on the surface with $\Delta E_{\text{vac}}$=0.61 eV for configurations $(a)$. For 4 vacancies per supercell, the surface configuration (i.e. $g$) yields the largest formation energy with $\Delta E_{\text{vac}}$=2.14 eV (see Table ~\ref{tab1}) and the most stable configuration correspond to vacancies all located inside the slab in configuration $(g')$ with a $\Delta E_{\text{vac}}$=1.36 eV (image 6 in Figure~\ref{fig1}). We stress that such a sudden change in relative stability as a function of vacancy concentration occurs already at zero field. In presence of the external field, the fully-emptied surface is further destabilized with respect to the fully-emptied middle chain (see Table ~\ref{tab1}). The energy difference between the two configurations, i.e. configurations $(g)$ and $(g')$ is 0.77 eV at zero field and becomes 0.91 eV at 10 V/nm. These results are consistent with a mechanism for O migration from the slab to the surface: as surface vacancies progressively increase with stronger applied fields, O atoms from the bulk migrate to the surface as a thermodynamically-driven effect.

\subsection{Electric field in presence of oxygen vacancies}
\begin{figure*}[ht]
\includegraphics[scale=0.4]{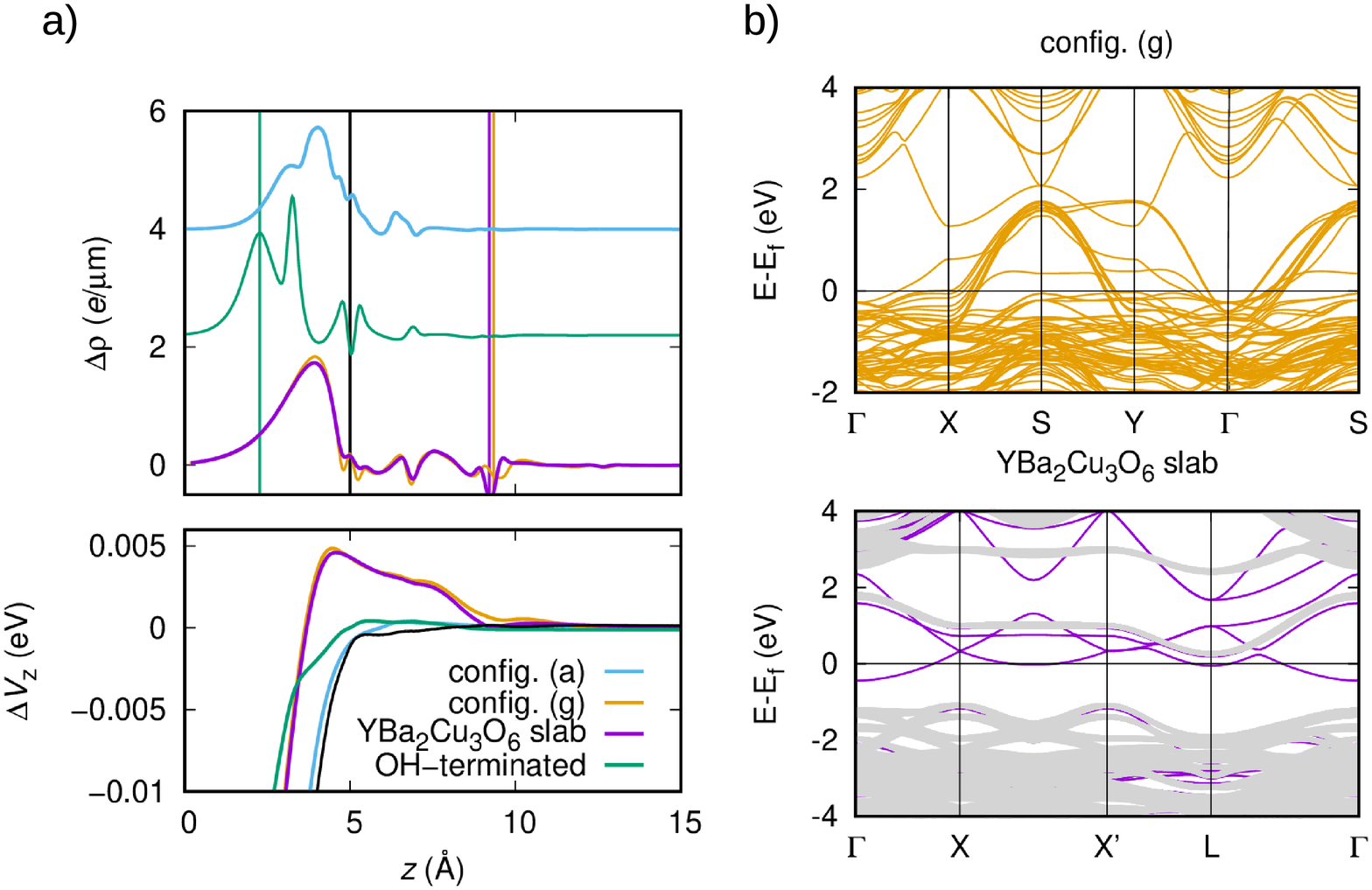}
  \begin{centering}
\caption
{a) $\Delta\rho$ and $\Delta V_z$ computed between 4 V/nm and zero field. The vertical black line indicates the the first plane of Cu atoms in all cases. The green line indicates the plane of H atoms used for the OH-terminated YBa$_2$Cu$_3$O$_7$ slab; the purple and orange vertical lines correspond to the position of the first CuO$_2$ in the corresponding case. b) surface-projected bulk band structure of YBa$_2$Cu$_3$O$_6$ (lower panel) and band structure of a YBa$_2$Cu$_3$O$_7$ slab (upper panel) terminated by planes of O-empty CuO chains in configuration $(g)$ at zero field. }
\label{fig4}
\end{centering}
\end{figure*}

We assess whether the electric field can penetrate into the slab when a progressively increasing amount of O vacancies are created, first at the surface and then within the chains throughout the slab. We employed a two-unit cell slabs along $z$ and either a primtive or a unit cell in-plane, depending on the case.
We computed the following scenarios: i) a slab with bulk stoichiometry YBa$_2$Cu$_3$O$_7$ and one vacancy per supercell on the surface, i.e. configuration $(a)$ in Figure~\ref{schema}; ii) a slab with YBa$_2$Cu$_3$O$_7$ bulk stoichiometry, and Cu-terminated, i.e. configuration $(g)$ in Figure~\ref{schema}; iii) a slab with stochiometry YBa$_2$Cu$_3$O$_6$ i.e. where all O atoms from every chain (there are three planes of chains in total) are removed; iv) a slab with YBa$_2$Cu$_3$O$_7$ bulk stoichiometry and OH terminated, i.e. a OH group is attached on top of each Cu at the surface.
As in the previous section, scenarios i), ii) and iv) were computed by performing a full geometrical optimization by fixing only the atomic positions of the surface plane far from the vacancies.
We recall that YBa$_2$Cu$_3$O$_6$ is a semiconductor exhibiting G-type antiferromagnetic order in the ground state \cite{Lopez2010,Poloni2018}. Thus, in absence of surface metallic states one would expect that the external field does penetrate into the slab. Scenario iii) was computed by adopting a $\sqrt{2}a\times\sqrt{2}a$ in-plane supercell in order to accomodate this antiferromagnetic phase. We perform then a full atomic relaxation of the bulk by imposing the G-type antiferromagnetic order and by fixing the lattice constant $a$ to the {\sl strained} value used so far. As discussed in our previous study\cite{Poloni2018}, a Hubbard-$U$ correction of 9 eV and 5 eV was used for Cu and inplane O atoms in order to recover the insulating phase. The slab is then built without allowing for further atomic relaxations.
Scenario iv) also represents a neutral slab but with a larger positive nominal charge on the Cu ion. If we assume Cu(I) for the ions on the chains, in presence of OH$^{-}$ these become Cu(II).
The planar averages of $\Delta V_z$ and $\Delta \rho$ computed for cases i)-iv) are reported in Figure~\ref{fig4} where all slabs are aligned to the first plane of Cu atoms (vertical black line at 5 \AA\ in upper left panel of Figure~\ref{fig4}). In none of the cases shown in Figure~\ref{fig4} does the electric field penetrate into the slab. While for YBa$_2$Cu$_3$O$_7$ the electric field is fully screened in correspondance of the CuO$_2$ plane (see black curve in the lower panel of Figure~\ref{fig4}), for i)-iv) the field is screened at the surface. Since the electrostatic potential profile along $z$ for a slab with two alternated superficial vacancies (i.e. configuration $(e)$) is similar to configuration $(a)$ it is here not shown for clarity.
For cases ii) and iii) a strong depolarization field is found within the first three atomic planes and it vanishes at the CuO$_2$ metallic plane (vertical orange and purple lines in upper panel a of Figure~\ref{fig4}). The reason behind this depolarization effect may be found in the free charge carriers at the surface in both situations. In the case of the YBa$_2$Cu$_3$O$_6$ slab, this can be understood by looking at the comparison between the surface-projected bulk band structure of YBa$_2$Cu$_3$O$_6$ and the electronic structure of the YBa$_2$Cu$_3$O$_6$ slab of Figure~\ref{fig4} (lower panel b). While bulk YBa$_2$Cu$_3$O$_6$ is a semiconductor, the corresponding Cu-terminated YBa$_2$Cu$_3$O$_6$ slab exhibits two metallic states on the surface that fully screen the electric field. In the case of YBa$_2$Cu$_3$O$_7$ with a fully emptied surface in configuration $(g)$ we see a similar situation. By comparing the band structure of configuration $(g)$ in Figure~\ref{fig4} with the case of stochiometric YBa$_2$Cu$_3$O$_7$ slab in Figure~\ref{fig2} we can identify the states arising from the surface defects. The state crossing the Fermi level along the $\Gamma$-X path both in the YBa$_2$Cu$_3$O$_6$ slab and in the superficially-emptied YBa$_2$Cu$_3$O$_7$ slab in Figure~\ref{fig4}) exhibit similar character and localize on the superficial Cu extending towards the apical O.

\begin{figure}[ht]
  \begin{centering}
  \includegraphics[scale=1.1]{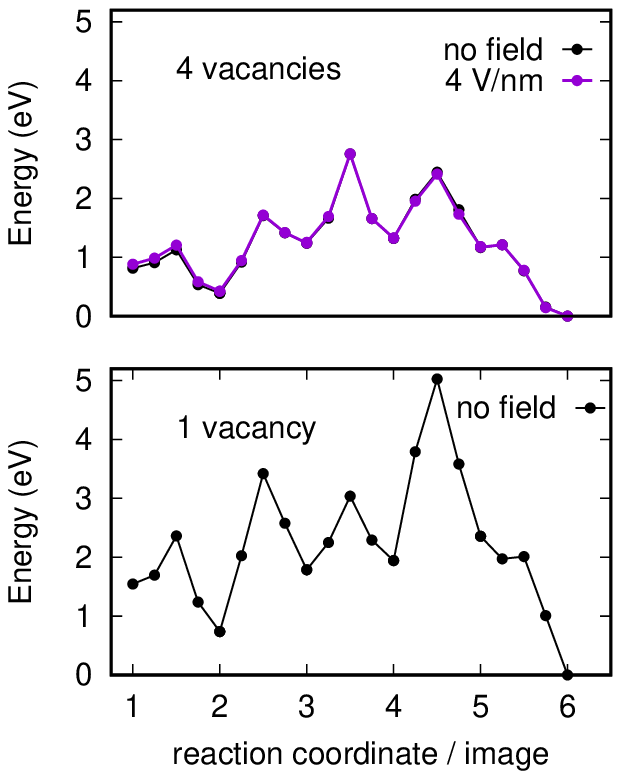}
\caption{Upper panel: potential energy profiles along the minimum energy pathways of O-vacancy diffusion across the slab, from the surface position (configuration $(g)$, image 1) to the center of the slab within the chain (configuration $(g')$, image 6), with and without external electric field. Lower panel: same thing for a O-vacancy diffusion starting from configuration $(a)$.}
\label{neb}
\end{centering}
\end{figure}

\subsection{Diffusion barriers}
 The minimum energy profile for the O-vacancy diffusion across the slab was computed by using the CI-NEB method \cite{PhysRevLett.72.1124}. Since bulk vacancies become energetically more favored than surface vacancies only at high concentrations (see configuration $(g')$ versus $(g)$), we compute the minimum energy path for oxygen diffusion across the slab, for all O atoms moving together from the surface towards the center of the slab. Practically, this is done by employing the in-plane primitive cell. This corresponds to a migration of all surface O atoms together in a concerted movement as imposed by the symmetry of the cell. The path shown in Figure~\ref{fig1} was computed using 5 images for each of the 5 sub-paths (see Figure~\ref{fig1}), i.e. between one O position and the next across the slab. The results shown in Figure~\ref{neb} show that the energy profile and the large diffusion barriers computed with no field are negligibly modified by the external field. A small change in presence of the field can be observed only for the first pathway, when the vacancy moves from the surface to the apical position, i.e. from image 1 to image 2, corresponding to configurations $(g)$ and $(h)$ in Figure~\ref{schema}, in agreement with our previous analysis of the fieldpenetration length.
 Since this concerted movement of all O atoms together imposed by the model may not represent correctly the true dyamics, the energy profile of one O-vacancy migration across the slab was computed starting from configuration $(a)$ in Figure~\ref{schema}, i.e. allowing only 1 vacancy per 2$\times$2 unit-cell. Another O-vacancy defect was kept fix in the chain in the middle of the slab throughout the diffusion pathway. In order to reduce the computational resources we allow only the O atom to move during the path optimization. The lower panel of Figure~\ref{neb} show again large barriers thus preventing us from using this mechanism to explain O diffusion across the slab.

\section{Conclusion}
We have studied the evolution of the electronic structure and the change in vacancy formation energy for a slab of YBCO as a function of an applied external field. We demonstrate that massive external fields induce isolated O vacancies at the surface. As the vacancies accumulate at the surface the defect configuration with vacancies in the center of the slab becomes preferred thus establishing a driving force for O migration towards the surface. These O atoms at the surface are in turn pulled out of the surface when external charges, like the polarized ionic liquid in the experiment, exert an electric force on the surface electrons. The relative stability of surface versus bulk vacancies change when defects accumulate, even at zero field, and the external field only slightly enhances this difference. From these calculations we conclude that the role of the external field appears to be limited to this electric force pulling charge outwards as the field never penetrates into the slab. Even when the bulk is fully emptied of O atoms from every chain, thus yielding a stoichiometry of YBa$_2$Cu$_3$O$_6$ which is semiconductor in bulk, surface states arise with metallic character and fully screen the external field.
More studies are required to identify the diffusion mechanism as the model employed in this work yields migration barriers too high to explain the experimental results.

\begin{acknowledgments}
Calculations were performed using resources granted by GENCI under the CINES grant number A0040907211. Additionally, the froggy platform of the CIMENT infrastructure was employed. The authors thank C. Attaccalite for discussions and for reading the manuscript. 
\end{acknowledgments}

\bibliography{./bibtex}

\end{document}